# Towards Noncommutative Topological Quantum Field Theory: New invariants for 3-manifolds

**Zois, I.P.**

TRSC / PPC, 9, Leontariou Street, GR 153 51, Pallini, Attika, Greece

and

The American College of Greece, Agia Paraskevi, GR 153 42 Athens, Greece

i.zois@dei.com.gr , i.zois@exeter.oxon.org , izois@acg.edu

**Abstract**. We define some new invariants for 3-manifolds using the space of taut codim-1 foliations along with various techniques from noncommutative geometry. These invariants originate from our attempt to generalise Topological Quantum Field Theories in the Noncommutative geometry / topology realm.

## 1. NCG as the underlying geometry for physical theories

There is accumulating evidence in various space-time dimensions that NCTQFT should provide a framework general enough to incorporate all 4 known interactions in nature:

*Case I (4 dimensions):* From the influential article due to G. 't Hooft on dimensional reduction in quantum gravity [1] where the holography principle was introduced, we know that quantum gravity should be a topological quantum field theory.

The argument is quite elegant: Given a (pseudo) Riemannian 4-manifold (M,g) we adopt the standard approach that the dynamical variable of the theory is the metric g and quantum field theory corresponds to computing the partition function. Thus one has to perform the path integral

$$Z(M) \sim \int Dg \exp(iS_M) \quad (1)$$

where $S_M$ denotes the Einstein-Hilbert action on M

$$S = (1/2) \int_M R \quad (2)$$

with R the scalar curvature and in (1) the path integral is meant to integrate over all metrics g on the manifold M. If one were able to perform this path integral over all metrics, then the result should depend only on the topological data of the initial Riemannian manifold M. The first experimental evidence for holography (dimensional reduction in quantum gravity) was provided by the GEO 600 experiment. This is an interferometric gravitational wave detector outside Hannover in Germany. C.J. Hogan in a series of articles [2] gave an explanation of the mysterious (300- 1400)Hz "noise" as arising from non-locality mechanisms due to holography.

Concerning the other 3 interactions (strong, weak and electromagnetic) we know from Connes' et al. work [3] that they can be incorporated in a noncommutative (nc) space arising as the Cartesian product of a 4-dim spin Riemannian manifold times a discrete space of metric dim 0 and KO-dim 6 mod 8. The nc space has a corresponding algebra of the form

$$\mathbf{C} + \mathbf{H} + M_3(\mathbf{C})$$

where **C** denotes complex numbers, **H** denotes the quaternions and $M_3(\mathbf{C})$ denotes 3X3 complex matrices.

[*Aside Note 1:* The total algebra becomes $M_2(\mathbf{H}) + M_4(\mathbf{C})$ if we incorporate the higher dimensional analogue of the Heisenberg commutation relations and the phenomenologically more accurate Pati-Salam gauge symmetry, see [15]].

Hence if we combine the two, gravity plus gauge theories of Yang-Mills type (for the electroweak and strong interactions) we believe that NCTQFT should provide the correct framework to describe all 4 known interactions in nature.

*Case II (higher dimensions, string/M-Theory):* We know that nc spaces arise both as extra (nc) toroidal compactifications of matrix models (Connes-Douglas-Schwarcz [4]) and from open strings when a gauge B-field is turned on (Seiberg-Witten [5]). In both cases, assuming space-time has either 4 dimensions or higher, gauge fields present are the origins of noncommutativity of the underlying geometry.

## 2. Invariants for 3-manifolds emerging from flat connections

According to the Atiyah definition of topological quantum field theory [6], one starts with a manifold with boundary and the correlation functions of the theory take values in some vector spaces associated to the boundary manifold. In the simplest case where the 4-manifold has no boundary, the correlators are just numerical (topological) invariants of the (bulk) manifold. Let us focus on dimension 4 which is the macroscopic space-time dimension and carries greater geometric interest. Obviously in this case the boundary of a 4-manifold say M, will be a 3-manifold, say N, namely $N = \partial M$. We fix SU(2) as our working Lie group. Given a 3-manifold N (assumed closed, oriented and connected) with fundamental group $\pi_1(N)$, the set

$$R(N) := \text{Hom}(\pi_1(N), SU(2)) / \text{ad}(SU(2))$$

consisting of equivalence classes of representations of the fundamental group $\pi_1(N)$ to SU(2) modulo conjugation *tends to be discrete*. For example if N is a (rational) homology 3-sphere, then R(N) has finite cardinality and the trivial representation is isolated.

There is a 1:1 correspondence between the elements of the set R(N) above and elements of the set

$$A(N) := \{\text{flat SU(2)-connections on N}\} / (\text{gauge equivalence}).$$

*The bijection is the holonomy of the flat connection.* The crucial observation is this: Although R(N) depends on the homotopy type of N, we can get topological invariants of N (namely invariants under homeomorphism) if we use the moduli space A(N). Depending on how one "decorates" or manipulates the (gauge classes of) flat SU(2)-connections in the set A(N) above, one can obtain the following:

*a. The low energy limit of the Jones-Witten invariant for N* (see [7]).

For any flat connection A (plus a Riemannian metric on N), one can form the twisted de Rham differential

$$d_A = d + A \quad (3)$$

twisted by the flat connection A; this is the exterior covariant derivative with respect to the connection A; it is a differential, namely its square equals zero due to the flatness of the connection. Then one can form the twisted Laplacian $\Delta$ in the standard way

$$= d_A d_A^* + d_A^* d_A \quad (4)$$

(using the metric to define first the Hodge star operator " $*$ " and then the adjoint operator $d_A^*$ in the usual way)

$$d_A^* = (-1)^{kn+n+1} * d_A * \quad (5)$$

(acting on k-forms on an n-manifold) and then compute the *Ray-Singer torsion* T(N,A) of the flat connection A as follows:

$$\log[T(N,A)] = (1/2) \sum_{i=0}^{3} (-1)^i i \, \zeta'(\Delta_{i,A}) \quad (6)$$

where $\Delta_{i,A}$ denotes the twisted Laplacian acting on i-forms and

$$\zeta'(\Delta_{i,A})(0) = (-d/ds) \zeta(\Delta_{i,A})|_{s=0} = \log D(\Delta_{i,A}) \quad (7)$$

is the $\zeta$-function regularised determinant of the twisted Laplace operator. Finally we add all Ray-Singer torsions (for all gauge classes of flat connections) and we get a topological invariant for the manifold N. Thus the finiteness of the set A(N) is crucial for the convergence (finiteness in fact) of the torsion sum.

[*Aside Note 2:* The Ray-Singer torsion is defined using the twisted Laplacian which involves the choice of a Riemannian metric; Under certain conditions-vanishing of twisted de Rham cohomology groups- it can be proved that the torsion is nonetheless independent of the metric].

Recall that the $\zeta$-function of the Laplace operator $\zeta(\Delta_i)$ is by definition (for complex s) equal to

$$\zeta(\Delta_i) = \sum \lambda_n^{-s}$$

where we sum over all non-negative eigenvalues $\lambda_n$.

*b. The Casson invariant.*

Take again G = SU(2) and pick a Hegaard splitting on N. Then, assuming that the set R(N) is regular (namely that the 1[st] twisted de Rham cohomology groups of N vanish for all flat connections), each (gauge class of) flat connection in the set A(N) acquires an orientation and then we take the difference between the number of positively minus the number of negatively oriented flat connections. Although the orientation depends on the Hegaard splitting, the difference behaves like an index and it is independent of the Hegaard splitting of N. The Casson invariant is the above difference of the number of positively minus the number of negatively oriented gauge classes of flat SU(2)-connections. Again the finiteness of both sets R(N) and A(N) is crucial for the definition of the Casson invariant.

*c. Floer homology*

Take again G = SU(2) and consider the Cherns-Simmons 3-form as a Lagrangian density on N. The corresponding action functional

$$S = (k/4\pi) \int_N tr[A \wedge dA + (2/3) A \wedge A \wedge A] \quad (8)$$

defines a Morse function on the space of irreducible connections. Its critical points are the flat connections.

Then each element of A(N) aquires a ''label" which is the Morse index of the critical point; in ordinary finite dimensional Morse theory this is equal to the number of negative eigenvalues of the Hessian. But the Hessian of the Chern-Simons function is unbounded below and we get infinity as Morse index for every critical point. So naive immitation of ordinary finite dimensional Morse theory techniques do not work.

Floer in [8] observed the following crucial fact: if we pick a Riemannian metric on N, then considering the non compact 4-manifold R X N along with its corresponding Riemannian metric, a continuous 1-parameter family of connections $A_t$ on N corresponds to a unique connection **A** on R X N; then, choosing the axial gauge (0th component of the connection vanishes), the gradient flow equation for the Chern-Simons function S on N corresponds to the instanton equation on the non compact 4-manifold R X N.

$$\partial_t A_t = *F_{At} \Leftrightarrow F_A^+ = 0 \quad (9)$$

Then consider the linearised instanton equation

$$d_A a = 0,$$
where a is a small perturbation. This operator is not elliptic; we perturb it to
$$D_A = - d_A^* + d_A^+$$
to make it elliptic. Then the finite integer Morse index for each critical point comes as the relative (with respect to the trivial flat connection) Fredholm index of the perturbed elliptic operator $D_A$. In this way the moduli space A(N) aquires a **Z**/8 grading and then we follow ideas from ordinary finite dimensional Morse theory: We define the Floer-Morse complex using as generators the critical points and the ''differential'' is essentially defined by the flow lines of the critical points. Taking the cohomology in the usual way we get the Floer homology groups of N. The Euler characteristic of the Floer-Morse complex equals twice the Casson invariant (see [9]). Again the finiteness of the set A(N) and R(N) is crucial in this construction.

All the above constructions depend crucially on the fact that the set A(N) (or equivalently the set R(N)) is finite. Unfortunately this is true for a rather small class of 3-manifolds (eg homology 3-spheres). It would be a big generalisation if we could find another moduli space which is finite for a larger class of 3-manifolds and if possibly for all 3-manifolds. The key observation is that in fact such a set exists: *It is the space of taut codim-1 foliations modulo coarse isotopy*. As proved by Gabai in [10], this set is finite for all 3-manifolds (closed, oriented and connected). The use of foliations has an additional advantage: it brings noncommutative geometric techniques on stage.

## 3. New invariants for 3-manifolds emerging from foliations

A codim-1 foliation F on a 3-manifold N is given by an codim-1 (hence dim 2) integrable subbundle F of the tangent bundle TN of N. Locally this can be defined by a 1-form ω satisfying
$$d = 0 \quad (10)$$
A foliation is called *topologically taut* if there exists a circle $S^1$ intersecting transversally all leaves. It is called *geometrically taut* if there exists a Riemannian metric for which all leaves are minimal surfaces (they have mean curvature zero). It is called *homologically taut* if there exists a closed 2-form β which is positive along the leaves. *These 3 different definitions of tautness are equivalent*. Two foliations are called *coarse isotopic* if up to isotopy of each one of them their oriented tangent planes differ pointwise by angles less than π. For any 3-manifold N (assumed closed, oriented and connected) we define the set G(N) as follows:

G(N) = {taut codim-1 foliations on N} / (coarse isotopy)

and let us denote by g(N) the cardinality of the set G(N). Then Gabai in [10] proved that g(N) is finite for all 3-manifolds (closed, oriented and connected) and it is a topological invariant of the 3-manifold N. The crucial fact is that although coarse isotopy depends on the choice of a Riemannian metric (since it involves the notion of angle), the number g(N) does not.

Armed with Gabai's result, there is the *possibility to define new invariants* and constructions on N by using different "labels" on elements of the set G(N).

  . *The Godbillon-Vey (GV for short) invariant of the manifold N*

Each codim-1 foliation F on the 3-manifold N has a 3-dim characteristic class which is the *Godbillon-Vey class of the foliation*. This is defined as follows: The integrability condition (10) is equivalent to d = for another 1-form θ. Then θ∧dθ is a 3-dim real cohomology class of N. We add all the GV classes for all elements in G(N). The result will be another 3-dim real cohomology class of N and then we can evaluate it on the fundamental homology class of N and thus get a real number as a result. This is the *GV invariant of the manifold N* (not the foliation).

*b. The sum of the Ray-Singer torsions for foliations*

A codim-1 foliation F on a 3-manifold N is given by an codim-1 (hence dim 2) integrable subbundle F of the tangent bundle TN of N. Integrability means that the Lie bracket of vector fields is closed in F. Equiavlently, this means that the horizontal (or tangential or leafwise) exterior derivative $d_F$ which is the usual exterior derivative restricted to take derivatives along the horizontal (or tangential or leafwise) directions only is a differential, namely $d_F^2 = 0$. We can use a Riemannian metric and define the Hodge star operator along with the adjoint horizontal differential $d_F^*$ and finally we can define the tangential Laplacian in the usual way (see section 2 above)

$$\Delta_F = d_F d_F^* + d_F^* d_F \quad (11)$$

Then we can define the Ray-Singer torsion for the foliation F in an analogous way as we did for flat connections (see equation 6 above)

$$\log[T(N,F)] = (1/2) \sum_{i=0}^{3} (-1)^i i \, \zeta'(\Delta_{i,F}) \quad (12)$$

with the same definition for the $\zeta$-function regularized determinant (equation 7). Finally we can add all Ray-Singer torsions for each foliation coarse isotopy class in G(N) and thus get an invariant for the 3- manifold N.

*c. Sum of Tangential Chern-Simons 3-forms*

Again F is a codim-1 foliation on the 3-manifold N. Since the tangential exterior derivative $d_F$ of the foliation F is a differential, we can form the tangential de Rham complex $(\Omega_F^\bullet(\ ), d_F)$, where $\Omega_F^\bullet$ denotes sections of the exterior bundle $\Lambda^\bullet(F^*)$. Taking the cohomology in the usual way we get the so-called *tangential cohomology groups* $H_F^\bullet(N)$. One can build the tangential version of the usual Chern-Weil theory of characteristic classes and finally construct the tangential Chern character from K-theory to tangential cohomology as a ring map

$$ch_F : K^0(N) \to \bigoplus_n H_F^{2n}(N;\mathbb{C})$$

From the above constructions it follows that tangential Chern classes have their corresponding tangential Chern-Simmons forms (see [11]). One can take the tangential Chern-Simons 3-forms for elements in G(N), add them up and integrate over the fundamental 3-dim homology class of N. This will give another numerical invariant for N.

*d. Transverse fundamental cyclic cocycles of foliations*

In general, given a codim-q foliation F on an n-manifold N, there is the noncommutative approach to the study of foliations. This amounts to defining the holonomy groupoid Gr(F,N) of the foliation and then complete it to a C*-algebra, this is the corresponding foliation C*-algebra denoted C(F,N). This construction is complicated and can be found in detail in Connes' book on Noncommutative Geometry [12]. Moreover, to each foliation F as above on N (with the mild assumption that its normal bundle Q: = TN/F is orientable) one can construct its *transverse fundamental cyclic cocycle* (see Zois [13]). This is a q-dim class (where q is the codimension of the foliation F) in the cyclic cohomology $HC^q(C^\infty(F,N))$.

[Aside *Note 3:* Actually in the construction of the tfcc it is more convenient to use the algebra of smooth complex functions $C^\infty(F,N)$ which is a dense subalgebra of C(F,N) because the cohomology of C(F,N) is rather poor in many cases].

In our case at hand, taut foliations by definition (see topological tautness above) have a closed transversal which is an $S^1$. If a foliation admits a closed transversal, then its corresponding C*-algebra simplifies drastically and it is reduced to just the algebra of continuous functions on the transversal.

[Aside *Note 4:* This is actually true in the case where the transverse circle intersects every leaf only once. We make this simplifying assumption here].

Hence in our case where we consider taut codim-1 foliations on a 3-manifold N, the foliation C*-algebra will just always be C($S^1$), ie the algebra of continuous complex valued functions on $S^1$ vanishing at infinity. Consequently for each taut foliation coarse isotopy class in the moduli space G(N), its corresponding transverse fundamental cyclic cocycle defines an element in $HC^1$ (C ($S^1$)), the first cyclic cohomology group of the algebra C ($S^1$). One can add all the corresponding transverse fundamental cyclic cocycles for elements on G(N) and get again as a result an element in $HC^1$ (C ($S^1$)). Thus we have to figure out what $HC^1$ (C ($S^1$)) is.

There is a theorem of Connes (see [12]) according to which (for compact M) there is a canonical isomorphism

$$HC^k(C(M)) \quad Z^{dR}_k(M) + H^{dR}_{k-2}(M) + H^{dR}_{k-4}(M) + \ldots,$$

where $Z^{dR}_k(M)$ denotes the set of closed de Rham k-currents, and $H^{dR}_j(M)$ denotes the ith de Rham homology group.

Hence in our case at hand we have $HC^1(C(S^1)) \quad Z^{dR}_1(S^1)$.

Since the degree is the same as the dimension, one can only integrate over $S^1$ itself and $Z^{dR}_1(S^1)$ must be 1-dimensional, generated by the fundamental homology class [$S^1$].

If the transverse circle does not intersect every leaf only once, then the foliation algebra will be more complicated. Nonetheless it will still be a simple C*-algebra though.

These new invariants need further study. In the generic case, where the assumption in Aside Note 4 above does not hold, what can go wrong is that the tfcc may be invariant under an equivalence relation which is more narrow than coarse homotopy and in fact each coarse homotopy class contains an infinite number of distinct tfcc. In this case convergence issues arise and one must be able to somehow control the infinite sum. (This is currently under investigation).

Looking further ahead, the possibility to define a noncommutative version of Floer Homology (see [14]) for all 3-manifolds using NCG tools and techniques starts from a Lagrangian density whose stationary points are taut codim-1 foliations. This is known as the "inverse problem" in the calculus of variations. Hopefully we shall be able to report on this elsewhere.

*Acknowledgements:* The author wishes to thank John Ball, Danny Calegari, Joachim Cuntz, Thierry Fack, Keith Hannabuss and Georges Skandalis for correspondence.